\documentclass{elsarticle}

\usepackage{amssymb}
\usepackage{color}
\usepackage{url}
\usepackage{hyperref} 
\usepackage{float}
\usepackage{setspace}
\usepackage{graphicx}
\usepackage{caption}
\usepackage{subcaption}

\begin{document}

\begin{frontmatter}



\title{A Conversational Approach to Well-being Awareness Creation and Behavioural Intention}


\author{Antonia Azzini\corref{cor1}}
\author{Ilaria Baroni}
\author{Irene Celino}
\address{Cefriel - viale Sarca 226, Milan, Italy,\\
E-mail: (antonia.azzini,ilaria.baroni,irene.celino)@cefriel.com}

\begin{abstract}
The promotion of a healthy lifestyle is one of the main drivers of an individual's overall physical and psycho-emotional well-being. Digital technologies are more and more adopted as ''facilitators'' for this goal, to raise awareness and solicit healthy lifestyle habits. 

This study aims to experiment the effects of the adoption of a digital conversational tool to influence awareness creation and behavioural change in the context of a well-being lifestyle. Our aim is to collect evidence of the aspects that must be taken into account when designing and implementing such tools in well-being promotion campaigns.

To this end, we created a conversational application for promoting well-being and healthy lifestyles, which presents relevant information and asks specific questions to its intended users within an interaction happening through a chat interface; the conversational tool presents itself as a well-being counsellor named Allegra and follows a coaching approach to structure the interaction with the user. 

In our user study, participants were asked to first interact with Allegra in one of three experimental conditions, corresponding to different conversational styles; then, they answered a questionnaire about their experience. The questionnaire items were related to intrinsic motivation factors as well as awareness creation and behavioural change. The collected data allowed us to assess the hypotheses of our model that put in connection those variables.

Our results confirm the positive effect of intrinsic motivation factors on both awareness creation and behavioural intention in the context of well-being and healthy lifestyle; on the other hand, we did not record any statistically significant effect of different language and communication styles on the outcomes. 

\end{abstract}


\end{frontmatter}

\section{Introduction}
\label{intro}

Attention to healthy lifestyles, on a physical and psychological level, is significantly increased in recent years, thanks also to the use of ICT technologies, that permeate various aspects of human life, becoming an essential part of it. Various digital tools are used every day, supporting people in carrying out different activities, and leading them to achieve greater awareness about the importance of a healthy lifestyle.

Change towards a better lifestyle is constantly promoted and advertised by the WHO\footnote{Cf.\url{https://www.who.int/}}, by emphasising the importance of both physical and psycho-emotional aspects of a healthy life. Changing perspective is the first step to increase the intention to implement changes that are aimed at personal improvement.

Digital supports for healthcare simplify the performing of various tasks, which can also be done remotely. People increasingly develop self-management in performing short and simple tasks, in the care of healthy lifestyles and well-being. Self-management~\cite{IJERPH22} refers to the set of relevant skills of individuals to manage and monitor their life. Self-management is at the core of interventions that can become very effective in changing behaviour of lifestyles and attitudes, as it helps individuals to increase awareness of their health through specific target information~\cite{JMIR21}. 

It is on such a basis that our work intends to make its own original contribution, to facilitate greater awareness and thus lead more individuals towards a more positive attitude to well-being. The first goal of our work is to delineate and better understand which factors contribute to a person's motivation to undertake and continue a lifestyle change. 

Among the digital tools, human-machine cooperation and interaction tools play an increasingly predominant role, and among them, particularly interesting are those based on conversational approaches for user involvement~\cite{Laban2019,ACMreview}. By simulating interpersonal interactions, such tools are mainly used to gather different types of information about users, as needed. For example, they can be used to obtain information on habits and lifestyles, but also on behavioural attitudes and perceptions, and to validate usability and trust in the tools themselves~\cite{kocaballi2022,Callejas2021,Tudor20}. Such tools also assume a significant role in supporting self-management practices.
Following this idea, we design and implement a conversational tool that acts as a counsellor to promote healthy habits and lifestyles. The aim is to gather information in order to identify which factors promote an increase in awareness and behavioural changes.

Moreover, in conversational system design, the choice of style with which the interaction with the user will take place, represents a critical aspect~\cite{Liebrecht2021}, since it significantly affects the user's interest and engagement in the tool itself. For this reason, in this study, we also aim at assessing, through the implementation of three different variants of the conversational tool, the effects produced by the use of different vocabularies and linguistic elements.

The remainder of this paper is organized as follows: we first present the theoretical bases on which this study is grounded in Section~\ref{sec:related}. In Section~\ref{sec:rq} we explain the hypotheses of our model and the research questions we intend to address. We then describe the conversational approach that we used to collect data and validate our model in Section~\ref{sec:method}, together with the details on the experimental protocol we followed. We illustrate our study results in Section~\ref{sec:results}: in the first part we present the exploratory investigation of the influencing factors and correlations among them; in the second part, we present the results from the confirmatory factor analysis and the final validated model. In this second part we also present the most relevant aspects resulting from a content analysis and we discuss the results of the sentiment analysis. Finally, we give an interpretation of the results and we close the paper with conclusions and next steps in Section~\ref{sec:concl}.

\section{Related work}
\label{sec:related}

This study considers various contributions of the literature referring to different aspects related to conversational agents, as examples of human-machine interactive solutions, their interaction styles, their application to health and well-being research fields, and, last but not least, their use to enhance awareness creation and soliciting healthy lifestyles in humans. The most relevant contributions, for the aim of this study, are reported as follows.

\subsection{Conversational agents and relevant factors}

In recent years, the growth of conversational technologies encourages humans to interact with conversational interfaces through various interaction ways and increasing communication channels to different application domains. Among these, conversational agents are taking on a crucial role in healthcare and prevention~\cite{Tudor20}. Several works presented examples of digital assistants that interact with humans as trainers, coaches, or simply as supports in the performance of activities aimed at improving lifestyle and, above all, human involvement.

The research field of Human-Computer Interactions (HCI) pays particular attention to the cognitive aspects that are useful to increase, for example, motivation, satisfaction and user engagement. The study reported in~\cite{Metux}, for example, presented different approaches for the overall user experience evaluation process. Among them, the authors referred, in particular, to the Self Determination Theory (SDT)~\cite{SDT}, which provides a mature and empirically validated approach to analyse the factors that promote sustained motivation and well-being. SDT identifies a small set of psychological needs considered essential for self-motivation and psychological well-being in individuals. Such needs can be summarised in user autonomy, competency and relationships: according to such a theory, ``if you increase autonomy, commitment will improve, if you increase competence, motivation will increase, and if you increase relationships, well-being will be improved". Therefore, the needs become valuable variables that should be evaluated in the design process of an HCI solution.

Following these aspects, an interesting work~\cite{ACMreview} reported a literature review about user experience evaluation on Conversational Agents (CAs) and their interactions with humans. The authors presented different contributions focused on how CAs may increase user engagement, enhance user experience, and enrich the relationships w.r.t. humans. The work also presented an interesting summary of metrics used in analysing a set of relevant factors influencing the overall user evaluation. The most valuable factors from such an analysis are user experience and behaviours, system usability, and perceptions. Many of these factors are part of the Intrinsic Motivation Inventory (IMI) Scale~\cite{IMI}, a well-defined multidimensional scale, initially developed for supporting the Self Determination Theory (SDT)~\cite{SDT}, and then used to measure the personal experiences of participants involved in a specific computer interaction task. The IMI scale has been broadly applied in the past, in particular, to define, among those considered, the most influencing metrics that can be used in a machine-interaction system optimization~\cite{IMItesting}.

Also Explanaible Artificial Intelligence (XAI) has contributed to such influencing factor analysis, by studying interesting aspects related to the human perspective in the definition of an HCI process: for example, in \cite{Liao2021} the authors highlighted the importance of providing a clear description and topic explanation in a given context, by considering the 4W questions (Who - Why - How - When) for detailing as more information as possible, while in~\cite{Liao2020} the authors set a question's bank used to define an explanation model based on AI. Another interesting aspect considered by XAI referred to the application of cognitive science and psychology to facilitate user understanding. In~\cite{roadmap}, the authors presented, in this direction, a deep analysis of cognitive states, and defined a set of metrics used to evaluate human factors. Examples of the summarised metrics are satisfaction, perceived goodness, attention, trust, and more technical ones like system interaction, ease of use and validation. Particular attention has been given to the user experience, that could be a critical aspect in interaction with AI, whenever less attention is given to human aspects such as satisfaction, understanding and user involvement, pillars of the user's trust\cite{Hsiao2021RoadmapOD}.

\subsection{Conversational agents for health and well-being}
In the health domain, CAs are playing an increasingly important role in physical and mental health care, bringing benefits, both to professionals or experts in healthy lifestyles and the users. In \cite{Brinkman2016} the authors focused the study from a human-centred point of view, based on the context and lifestyle definition, to better understand current and future users of CA-based solutions. The study indicated how such virtual health coaches may highlight important capabilities as 24/7 accessibility, data acquisition and management, consistency and perseverance. The reported advantages mainly referred to the acquisition of greater knowledge in the field and an increased user perception about his/her awareness. The users also gave particular importance to feeling they are not alone in their own devices in managing lifestyle changes, often not easy to undertake. Then, the results indicate a growing interest of users in solutions aimed at health prevention and lifestyle change, also thanks to the increased use of digital tools, even by older adults. According to this last aspect, more attention should be given to the different communication styles concerning the user. In another well-being promotion study \cite {dingler2021}, the authors report, for example, a review of recent advances in technologies underpinning conversational agents. In the study, they indicated how conversational agents have become valuable solutions in therapies to assist patients' self-management, influencing their behaviour and awareness, while at the same time training and educating them in healthier lifestyles.

A more recent summary of other relevant contributions has been presented in~\cite{kocaballi2022}, that highlighted user acceptance of CAs in the healthcare domain, together with an early promise in boosting healthcare outcomes in physical and mental health, and in well-being lifestyle. The work also focused on important aspects of such digital tools, such as the attention to the different responses given by different users in healthcare treatments, the use of suitable solutions to better understand the healthy status of interacting people, and the use of standards or well-known models to evaluate intrinsic human aspects as enjoyment, interest, and perceived usefulness.

Finally, in~\cite{Tudor20} the authors presented an extensive review of several contributions about health and well-being topics: the performed evaluation is relevant for application areas, such as education, health and well-being monitoring, and lifestyle behavioural changes. The contributions have been evaluated according to different performance metrics, such as accuracy, effectiveness and acceptability: an open issue emerging from such analyses focuses on assessing feasibility, acceptability, safety, and effectiveness of different CAs formats and styles concerning the target people’s needs and expectations. In this direction, intrinsic measurements could be further investigated.

\subsection{Conversational style for interactive agents}

An important aspect of human-machine solutions has been presented in recent research studies~\cite{Asbjorn19,Laban2019}, and it focuses on the importance of defining the most appropriate communication style in human-machine interaction. The work presented in~\cite{Liebrecht2021} analysed, for instance, the effects of conversational agents using an (in)formal communication style towards humans, the effectiveness of different types of interaction, and the various behaviours assumed by people. The paper also reported an interesting comparison of the approaches of two different communication styles (formal and informal), implemented with diverse linguistic elements.

Another interesting work, reported in~\cite{Tudor20}, presented an overview of different communication styles used in the literature, by highlighting what the authors call ``personality traits" of a conversational agent. Among them, particularly relevant are (1) coaching mode, defined through a motivating and encouraging style, (2) factual, defined through a non-judgmental approach, no personal opinions, and responses based on facts or observations, (3) knowledgeable, based on content created for a particular aim and (4) informal, which, like talking to a friend, uses exclamations, abbreviations, and emoticons. Regarding this last communication style, Fadhil and colleagues presented in~\cite{Fadhil2018} a study about the effects of emoji in a conversational interface-assisted health coaching system. The work explored the decorative use of emojis in both mental and physical well-being dialogues. Results indicated that the effects were directly related to the different health aspects: indeed, participants appeared confident in sharing information with the CA about their mental well-being when the dialogue included emojis, while they were less confident when the topic was about mental health. As a final remark, the authors indicated that communication with such digital tools is evolving also beyond simple text by adding content to the conversation, and by including photos, animated GIFs, location images, and so on. 

A similar result was reported in~\cite{CUI2023}, where the authors presented how the conversational style plays a significant role in creating positive user experiences supporting engagement, through the implementation of a virtual coach. The study presented a language kept simple and clear to avoid confusion and to allow an informal and colloquial tone for easy relations. The virtual coach used emojis to enhance the message, and, it was always designed to appear as friendly and empathic w.r.t. the final user. This style was adopted to deal with style problems reported in cases where communication styles appear robotic, i.e. not similar to human communication, and messages seem judgmental towards the user.

Finally, the work conducted in~\cite{Silva22} proposed guidelines for chatbot design. The study was based on a systematic literature review focused on conversational practices and their impacts on users. These guidelines were shared with practitioners to gather their impressions about the digital solution and evaluate its usefulness and ease of use.
The authors defined a design map for a chatbot-based conversational agent, taking into account the aim and impact on the final user. A diagram summarised a set of best practices to be considered in the design of a user-centred conversational system, together with a set of indications on actions that should be avoided. Critical aspects of such digital interaction solutions are, for example, related to the importance of providing transparency, clear information, trust and user engagement. The use of chatbots can become critical if it is not defined that one is talking to a digital system. Moreover, as reported in~\cite{luger2016like}, purely automated chatbots can become unnatural from the interaction point of view: the authors specify that, generally, it happens when conversational flow appears too linear, by introducing a gap between the user's expectation and the experience. 

\subsection{Conversational agents for behavioural change and awareness creation}

Behavioural change to adopt more healthy lifestyles is known as a complex process that requires time and effort. The literature shows how the study of behavioural change has led to the definition of various theoretical models, that highlighted the importance of both cognitive and context-related aspects for each individual~\cite{liao2021Survey,Metux,Miller17}. Particular attention was given to the explanation of content, which people generally appreciate if maintained  simple. Describing the context is important, but people find more engaging those solutions that contain the necessary information clearly and simply explained. For this reason, the position shared by the authors is that effective health promotion interventions must take these factors into account, not only for a social group but mainly for the individual, to increase awareness of the benefits that even small changes to his/her habits can produce.

The motivating activity is well described in the literature by the model related to \emph{intrinsic motivation} and self-regulation, often used to measure an individual's subjective experience. As stated above, the Intrinsic Motivation Inventory (IMI)~\cite{IMI} is a multidimensional scale developed on such an intrinsic motivation model, in support of the Self-Determination Theory (SDT)~\cite{SDT}. SDT is a validated theory that argues that motivation and regulation are driven by three innate needs: autonomy, belonging and competence. The IMI scale is defined by a series of constructs that highlight aspects such as interest, usefulness, value, perceived effort, and others that contribute to defining, for each user, the needs indicated in the SDT. In an online coaching environment, the IMI, allows, in the design phase, to highlight aspects related to the training of motivational styles, and their respective indications, enhancing the user's perception of competence, willingness to perform a task, and increasing self-awareness of his/her capabilities.

Other interesting actions suggested by the literature, at both contextual and cognitive levels, refer, for example, to different activities, such as upgrading context information and promoting decision-making processes. All of them aim at raising awareness on the consequences of specific behaviours, by proposing and stimulating virtuous behaviour, through the use of positive reinforcement, as suggested by the so-called Nudge theory~\cite{nudge2008}. As reported in \cite{Caraban19}, the concept of nudging indicates that there is knowledge of systematic biases in decision-making and that, such knowledge, can be exploited in a process to support people in making optimal choices. The study reported different nudging solutions, and, it presented, among them, the so-called "spark" nudges, useful for motivating and engaging users to change their behaviours.

In~\cite{Cecily2022}, relevant considerations about human factors also arise from research about the conversational agent's design. The authors compared two different ways to implement a digital tool, in order to assess differences in human interactions concerning motivation and behavioural change intention. The work was conducted in the context of a study in the medical field, specifically for campaigns against smoking. The comparative work between a motivational interviewing approach w.r.t. a counterfactual counselling approach showed useful suggestions for awareness and behavioural change definition.  

In this direction, other interesting aspects arise from the use of coaching models for the design of such digital tools design, focusing on the final user and his/her needs. An example in the literature is represented by the GROW model~\cite{grow}, which consists of four main steps: identify a clear and well-defined goal, analyze the actual context, identify choices and possible actions for a user in that context, define a plan and actual actions that the user can carry out to work towards awareness and intention to proceed with a specific change. The work carried out in \cite{Grow2020} shows an example, where the authors proposed a solution for eliciting behavioural changes by applying and extending the GROW model in the health domain.

An interesting work about the potential use of conversational interfaces for the specific area of coaching is presented in~\cite{Callejas2021}: in their contribution, the authors refer to recent literature about using conversational agents for soliciting healthy behavioural changes and awareness in the users. Moreover, they also cite solutions for enhancing user's self-reflection, for example by developing coaching and informative approaches. Also the already reported study \cite{Brinkman2016} focused the attention on emerging solutions based on virtual health agents able to support people in changing their behaviour. The analysis showed an interesting comparison of different skills, human needs and potential solutions that could be considered for improving human-machine interaction. The analysis also considered influence, attitudes and the resulting behaviour of people. Finally, it presented recent challenges concerning these aspects and highlights, for each of them, potentials and critical issues.

\section{Research Questions and Hypotheses}
\label{sec:rq}

In this paper, we aim to define a  model to assess the different factors influencing intentions to change behaviour for a healthy lifestyle, at the same time by increasing awareness in well-being, through a counselling conversational agent. In detail, with our approach, we intend to address the following research questions:

\begin{itemize}
    \item\textbf{RQ1}: can a (digital) conversational approach influence in creating awareness and soliciting a behavioural intention?
    \item\textbf{RQ2}: what is the influence of different motivation factors (interest, trust, usefulness, etc.) on awareness creation and behavioural intention?
    \item\textbf{RQ3}: how the conversational style (informal, informal with multimedia elements, and formal) can influence the factors in the model?
    \item\textbf{RQ4}: are there any differences in the participants' behaviour according to the topic of the conversation?
\end{itemize}  

In order to answer RQ1, we focus on two main metrics, respectively, \emph{behavioural intention}, which represents the individual perceived probability that he/she will use the conversational tool in the future (often used as main dependent variable in the TAM model~\cite{TAM}), and \emph{awareness creation}, which represents the individual perceived ability to learn about and reflect on a (possibly new) topic, internalising relevant information. Furthermore, we also perform a sentiment analysis for a more qualitative assessment of the participants' reaction to the conversation.

To answer RQ2, we focus on some relevant factors from the IMI Scale~\cite{IMI}. Such a scale is defined on an adopted model related to \emph{intrinsic motivation} and self-regulation, and it is often used to measure an individual's subjective experience. 
We analysed all IMI constructs, together with their descriptions, and selected those that highlight aspects of the user experience that are closely related to the person's awareness and behavioural change.  
The most fitting constructs for our study are: \emph{interest}, that provides a measure of how enjoyable and fun the individual considers an activity to be (in our study, activity refers to interaction with the conversational counsellor); \emph{value}, that indicates how important the individual rates the activity, and how useful this activity is perceived to be; \emph{trust}, a measure of how reliable and confident the individual considers the tool, in order to perform the task, or, more generally, to achieve a goal; and \emph{relatedness}, that is the measure of how an individual places his/herself in a relationship, or an interaction with an external subject (in our case, the conversational tool). 
Our goal is to investigate the influence of intrinsic motivation constructs on the previous two dimensions, behavioural intention and awareness creation.

\begin{figure}
    \centering
    \includegraphics[width=0.7\textwidth]{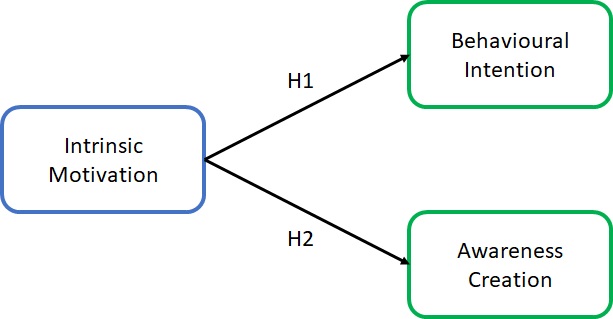}
    \caption{Representative schema of the defined model}
    \label{fig:model}
\end{figure}

Therefore, the research hypothesis for the validation of the model are as follows (cf. also Figure~\ref{fig:model}).

\begin{itemize}
    \item \textbf{H1}: the \emph{intrinsic motivation factors} (interest, value, trust, relatedness) influence the \emph{behavioural intention}
    \item \textbf{H2}: the \emph{intrinsic motivation factors} (interest, value, trust, relatedness) influence the \emph{awareness creation}
\end{itemize}

To meet RQ3, we introduce the \emph{conversational style} as an experimental condition to perform an A/B testing and we measure if this factor has any influence on users' behavioural intention. More details on this aspect are offered in the next section.

To meet RQ4, we perform an analysis of the choices of the participants in the conversation concerning the following topics: \emph{healthy eating}, \emph{physical activity}, \emph{social relationships} and \emph{mental health}.

\section{Methodology}\label{sec:method}

In this study, we adopt Coney~\cite{celino2020coney}, a digital toolkit to create (fully scripted, without AI support) conversational experiences. In this case, the user is invited to take part in a healthy lifestyle conversation, guided and supported, step by step, by Allegra, our well-being counsellor. Each experience is implemented through a conversational agent, structured as a survey, that follows the GROW coaching model \cite{grow}. The survey is then designed according to the following four steps: 
\begin{itemize}
    \item \emph{Goal}: the user is supported to define a specific goal  in the healthy lifestyle context;
    \item \emph{Reality}: the user is asked about their actual lifestyle and habits;
    \item \emph{Option}: the user is given a set of different ideas with the goal of improving their lifestyle;
    \item \emph{Will}: the user is solicited to express their intention to carry out an action plan to execute a selected option to achieve their goal.
\end{itemize}

Table \ref{tab:growexample} shows an example of the application of the GROW coaching model into the Allegra counsellour, by showing, for each item of the model, a representative extract of the interaction with Allegra. 

\begin{table}[ht]
    \centering
    \caption{Example of application of GROW model}
    \label{tab:growexample}
    \small
    \begin{tabular}{p{2cm}|p{9cm}}
\textbf{GROW Step} & \textbf{Example of interaction with Allegra counsellour} \\
  \hline
{Goal}  & I'm a well-being counsellor and, if you like, I'd be glad to discuss with you a healthy lifestyle topic: psychological and emotional well-being: how about it?\\
{Reality}  & Have you actively taken care of your psychological and emotional well-being lately?\\
{Option}  & So, about psycho-emotional well-being, which of these topics is more important to you?\\
{Will}  & Can we now focus on developing some action plan for you?\\
\end{tabular}
\end{table}

The conversational experience covers some main aspects of the psychological and emotional well-being in a healthy lifestyle, corresponding, respectively to \emph{self-care}, \emph{social relationships}, \emph{mental health} and \emph{healthy eating}. The interaction with the user is based on informative messages, assertions, rhetorical questions or specific requests to the user, to gain information about their reality, options, preferences, and personal intentions. The conversation allows the user the option to choose between two predefined macro domains: "Mental health and healthy eating" and "Social care and relationship". Consequently, the content of his interaction with the conversational agent will vary according to this choice initially taken. Finally, to assess the user intention to change his/her behaviour aided to a more healthy lifestyle, the survey asks the user about his/her willingness to follow a well-being related suggestion in the following month.

To evaluate the well-being awareness creation and the influencing factors for soliciting a behavioural change, we compared three conversational experiences implementing three different experimental conditions of the human-machine interaction process with the well-being counsellor. Each experience differs from the others in its conversational style (cf. RQ3), according to the use of different linguistic elements, respectively\footnote{The interested reader can try out the previews of the three experimental conditions of the Allegra well-being counsellor at the provided web links.}: 
\begin{itemize}
    \item \emph{\href{https://coney.cefriel.com/app/chat/?d=T6QexsICi1QxzG21-ZDakZB8S-GJkgDUvCU-I3quphZUu1IbVxJ-nw0tbSV-d5DrkAoodvBRxJfuh8gIkC-vkA==}{Informal style}}, with sound mimicking, contractions and shortenings, active voice, informal vocabulary, etc.;
    \item \emph{\href{https://coney.cefriel.com/app/chat/?d=T6QexsICi1QxzG21-ZDakaS5rKA4G0U46wota0nrAhtAfbJwaRijcfqX3asIgpHT1GbYtSL6BLeVQM6HjA5DjA==}{Informal style with multimedia elements}}, an extended version of the previous informal one, with emoticons and animated GIFs; and 
    \item \emph{\href{https://coney.cefriel.com/app/chat/?d=T6QexsICi1QxzG21-ZDakWunopPQUbZz8Q2_4eMomrbq13yb6Y8vrFrKqO5b2laDRDZ3nHpbjcg4U3ZvvhC90A==}{Formal style}}, with passive voice, formal vocabulary, and without all the verbal and non-verbal cues used for the informal versions.
\end{itemize}

\begin{figure*}[h!]
\centering
\begin{subfigure}{0.3\textwidth}
  \centering
  \includegraphics[width=\textwidth]{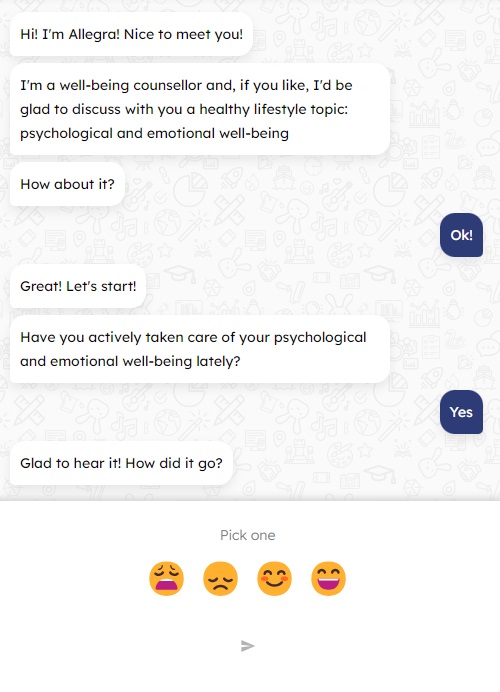}
  \caption{Informal}
  \label{fig:informal}
\end{subfigure}
\begin{subfigure}{0.3\textwidth}
  \centering
  \includegraphics[width=\textwidth]{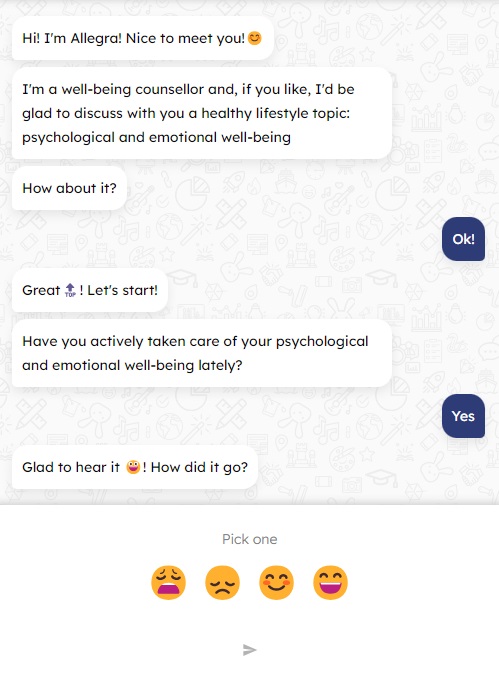}
  \caption{Informal with emoticons}
  \label{fig:informalgif}
\end{subfigure}
\begin{subfigure}{0.3\textwidth}
  \centering
  \includegraphics[width=\textwidth]{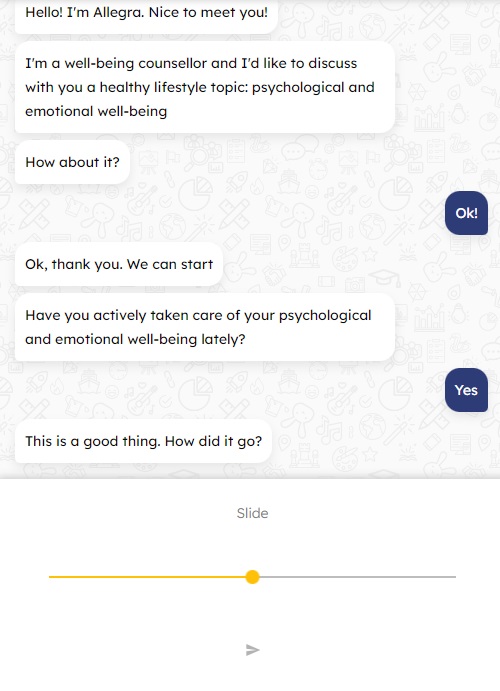}
  \caption{Formal}
  \label{fig:formal}
\end{subfigure}
\caption{Example of the screenshots extracted from the implemented conversations}
\label{fig:screenshots}
\end{figure*}

An example of the three conversational styles is reported in Figure \ref{fig:screenshots}.

After the interaction with the conversational tool, the user is provided with an assessment questionnaire, to evaluate their experience. 
The questionnaire is composed of a list of statements that the user is asked to rate on a 1-to-7 Scale (1 = not at all true, 7 = very true) for the assessment of all the constructs defined in our model. 
In particular, the questionnaire includes 3 items for each of the constructs. Table~\ref{tab:metrics} provides the full list of questionnaire items. The pre-testing of the questionnaire confirmed its reliability and validity as an assessment instrument.

\begin{table}[!ht]
    \centering
    \caption{Assessment evaluation questionnaire}
    \label{tab:metrics}
    \footnotesize
    \begin{tabular}{p{3cm} p{7.7cm}}
\textbf{Factor} & \textbf{Questionnaire item} \\
  \hline
{Interest}  & I would describe Allegra as very interesting\\
{Interest}   & Allegra did not hold my attention at all\\
{Interest}   & While using Allegra, I thought about how much I enjoyed it\\
{Value}   & I think Allegra is important because it can help me in improving  my personal lifestyle\\
{Value}    & I think that doing this chat with Allegra is useful for a healthy lifestyle\\
{Value}   & I would be willing to chat with Allegra again because it has some value to me\\
{Trust}   & I don't feel like I could really trust Allegra\\
{Trust}   & I think that chatting with Allegra is an important activity\\
{Trust} & I feel that Allegra is a trustable source\\
{Relatedness}   & I feel close to Allegra\\
{Relatedness}   & I'd like a chance to interact with Allegra again\\
{Relatedness}   & I'd really prefer not to interact with Allegra in the future\\
{Awareness Creation}  & Having a conversation with Allegra helped me on reflect on the topic\\
{Awareness Creation}  & Allegra helped me to become more aware about the importance of healthy lifestyle\\
{Awareness Creation}   & I don't think that my knowledge on healthy lifestyle can improve by chatting  with Allegra\\
{Behavioural Intention}  & I am confident that I will follow the Allegra's suggestion\\
{Behavioural Intention}   & I would be willing, with Allegra's support, to start changing my daily activities to improve my lifestyle\\
{Behavioural Intention} & When Allegra asked you about your willingness to follow a well-being related suggestion in the next month, what did you answer?\\
 %\hline
    \end{tabular}
\end{table}

For the validation of the model of Figure~\ref{fig:model}, we set up three crowdsourcing campaigns on the Prolific platform\footnote{Cf. \url{https://app.prolific.co}}~\cite{palan2018}, one for each of the implemented experimental conditions, involving a total of 600 users (200 participants for each campaign); the experiment was run during August 2023. 

For the selection of the crowd workers in each campaign, the inclusion criteria were: age between 20 and 60, residence in Europe, fluency in English, while, in order to implement an A/B testing, the exclusion criterion was the participation to the other two experimental campaigns. No inclusion or exclusion criteria were imposed on any other personal characteristic (gender, nationality, etc.). In each of the campaigns, the participants were introduced to the study context and were invited (1) to try the conversational experience discussing the healthy lifestyle topic with the Allegra well-being conversational tool and (2) to fill in the questionnaire to evaluate such experience, as explained above.

It is important to underline that no incentive mechanisms were used. The first reason is that recruitment on crowdsourcing platforms already includes economic compensation. Furthermore, our study wants to collect information without adding any bias to the naive behavioural intention of the final users. 

\section{Experimental Results}\label{sec:results}

In this section, we describe and discuss the experimental results we obtained applying the methodology explained above. We first present some general statistics on the experiments, then we discuss the outcomes we obtained w.r.t. our initial research questions, and finally we add further considerations on the topic choice and sentiment towards the experience.

\subsection{General statistics}

A first statistical analysis carried out on the three groups of participants, defined for the three experimental campaigns, shows a homogeneous and balanced distribution, in terms of gender, age, education and employment status. The data indicate that about 63\% of them are women and 37\% are men, with a quite balanced age distribution (23\% between 20 and 30 years old, 36\% between 31 and 40, 22\% between 41 and 40, and 19\% between 51 and 60). Regarding employment status, about 56\% of the participants declare to be full-time or part-time employed and only 10\% are students. 

The average compilation time of both Allegra conversation and evaluation questionnaire for all the participants was equal to 6.49 mins, and standard deviation of 3.23 mins. All the questionnaires items were checked by calculating the Cronbach alpha~\cite{brown2002cronbach}, showing high internal coherence of all the items for intrinsic motivation (0.95), awareness creation (0.81), and behavioural intention (0.81). Moreover, checking the individual items of the questionnaire, we noticed a high internal correlation, in particular among all intrinsic motivation items (0.95), which is another sign of reliability and consistency of our evaluation questionnaire. For this reason, as well as for the results illustrated in the following, we consider interest, value, trust and relatedness together as parts of a broader intrinsic motivation factor, instead of considering them separately.

\subsection{Ability of the conversational approach}

Our first research question aims at testing if a conversational approach is able in creating awareness and soliciting a behavioural intention. While of course, we could not measure the actual behavioural change of the involved participants, in order to give an answer to \textbf{RQ1} we analysed the scores given on the awareness creation and behavioural change items of the questionnaire.

\begin{figure}[ht]
    \centering
    \includegraphics[width=0.7\textwidth]{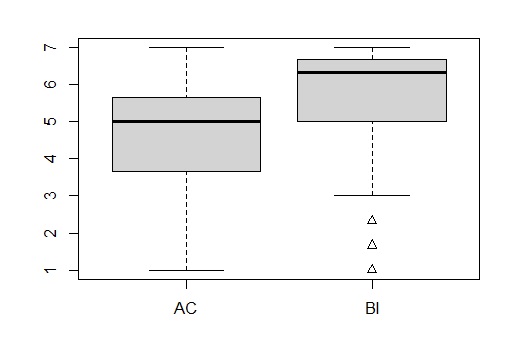}
    \caption{Boxplot of Awareness Creation and Behavioral Intention of the participants}
    \label{fig:boxplotAWABI}
\end{figure}

As depicted in Figure~\ref{fig:boxplotAWABI}, the data shows relatively high awareness creation (AC) (mean 4.58, median 5, std dev 1.37), and high behavioural intention (BI)(mean 5.79, median 6.33, std dev 1.23), on a scale from 1 to 7. This result induces us to confirm our hypothesis that indeed the conversational approach seems to be able in creating awareness and soliciting a behavioural intention.

\subsection{Sentiment analysis}

At the end of the questionnaire, we allowed the participants to freely comment about their experience, highlighting the best and most critical aspects of Allegra according to their opinions. 
Then, we manually performed a sentiment analysis on all the open comments by ranking them on a scale defined by 5 values (negative, quite negative, neutral, quite positive and positive). 
Each comment reported under "positive aspects" (pros) is evaluated and compared with the corresponding comment under "negative aspects" (cons) provided by the same user. A different rating score is then assigned, according to the following conditions:

\begin{itemize}
    \item\textbf{Negative}: it is assigned to both cases where the 'pros' do not highlight a user's opinion, but only features of Allegra, and the "cons" only report opinions expressing strong limitations and criticalities, and that do not help improve the solution.
    \item\textbf{Quite negative}: it is assigned to cases where the participants indicated limitations and disadvantages for the solution, with only marginal and not effective "pros". This rating is also assigned in cases of opinions that, although very critical and negative, they could become insights for improvements.
    \item\textbf{Neutral}: it covers those cases where users did not express an opinion, either positive or negative, but only reported features of Allegra. The same rating is also assigned to all cases where the participants preferred not to express their opinion.
    \item\textbf{Quite positive}: it is assigned to all cases where, although the participants observed some limitations and disadvantages of the solution, they mainly reported descriptions of positive experiences of using Allegra. This rating is also assigned to those cases in which negative opinions point out useful indications for future improvements.
    \item\textbf{Positive}: it is assigned to both the cases where the participants mainly reported positive advices, and also to those where, even apparently critical, participants are confident in the solution and provide very useful and constructive feedback.
\end{itemize}

The manual analysis of the sentiment was performed by the three authors separately and then aggregated by majority voting,  "translating" it into one of the five rating scores above mentioned. An example of the evaluation coming from the sentiment analysis, together with different examples of quotations is reported in Table~\ref{tab:SA}. In general, participants positively rated the experience with the digital counsellor (34\%  fully positive and 19\% quite positive out of the total 600 participants),  compared to a minority of negative sentiment respondents (14\% negative and 17\% quite negative) and only  16\%  neutral or ambiguous. 

\begin{figure}[H]
    \centering
    \includegraphics[width=1\linewidth]{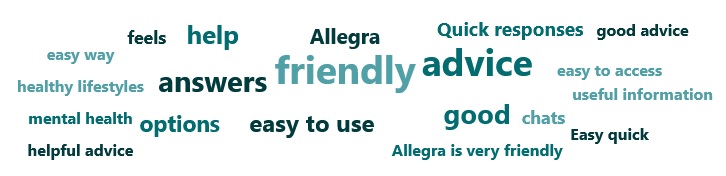}
    \caption{Word cloud of the most frequent answers from free-text comments of participants about the conversational counsellor}
    \label{fig:commenti}
\end{figure}

While we collected very wide and diverse opinions, it is worth noting that a large number of users commented that the tool was easy to use, friendly, immediate, interactive, informal, familiar, concise and informative; Allegra was perceived as non-intimidating, positive, encouraging, non-judgmental, different from a robot-style; the provided information and suggestions were considered relevant, helpful, interesting and useful, in some cases completely new. 
On the more critical side, some users found the suggestions too generic, the conversation a bit superficial, including information they already knew, and a few lamented the limited room for personalised advice (which actually was out of scope for our study). The positive comments outweighed the critical ones, confirming that the overall conversational approach was perceived as a pleasant experience (a word cloud from participants' answers is displayed in Figure~\ref{fig:commenti}). 

Finally, we analysed the possible effect of \emph{awareness creation} and \emph{behavioural intention} on sentiment analysis. We used a two-way ANOVA, which revealed that both \emph{behavioural intention} (F(1, 596)=65.16, p=0.0017) and \emph{awareness creation} (F(1, 596)=9.86, $p < 2.2e-16$) showed a statistically significant effect on the sentiment score; this means that there is indeed a relation between the actual intention to change behaviour and the acquired awareness with the sentiment towards the coaching experience. However, on the other hand,
there was no statistically significant interaction between the joint effects of \emph{behavioural intention} and \emph{awareness creation} (F(1, 596)=0.093, p=0.76).

\begin{table}[!ht]
    \centering
    \caption{Sentiment analysis: example of results and quotations}
    \label{tab:SA}
    \footnotesize
    \begin{tabular}{p{2cm} |p{1cm}| p{2cm} |p{5cm}}
\textbf{Label} & \textbf{Perc.} & \textbf{Description} & \textbf{Example quotation} \\
  \hline
{Positive}  & 19\% & A positive perception & Keeping the conversation interactive rather than just telling you what to do and also being realistic about whether you will keep up with goals. \\
{Quite Positive}  & 34\% & A mostly positive perception & Maybe an opportunity to reflect on different aspects of health and well-being than one would individually. \\
{Neutral}  & 16\% & An ambiguous, not clear perception & Not sure I just enjoyed the chat, staying neutral. \\
{Quite Negative}  & 17\% & A mostly negative perception & It feels like it is giving canned answers. I would prefer to speak with a human face to face. \\
{Negative}  & 14\% & A negative perception & Too generalised, very standard advice, not able to adapt to human difference. \\
 %\hline
    \end{tabular}
\end{table}

\subsection{Influence of intrinsic motivation}

Our second research question aims at evaluating the effect of intrinsic motivation on awareness creation and behavioural intention. To answer \textbf{RQ2}, we tested both the correlation and the causal influence (through structural equation modelling) of intrinsic motivation factors over our two main outcome factors.

As shown in Table~\ref{tab1}, all the intrinsic motivation items (interest (INT), value (VAL), relatedness (REL) and trust (TRU)) have a moderate positive correlation with behavioural intention (BI) items (values ranging from 0.54 and 0.58) and a strong positive correlation with awareness creation (AC) (values ranging from 0.73 to 0.81). This is the first result that suggests an actual positive effect of intrinsic motivation (IM).

\begin{table}[ht]
\centering
\caption{Correlations among the item groups}\label{tab1}
\begin{tabular}{p{1cm} c p{1cm} p{1cm} p{1cm} p{1cm} c p{1cm}  }
 && \textbf{INT} & \textbf{VAL} & \textbf{REL} & \textbf{TRU} && \textbf{IM} \\
\hline
\textbf{AC} && 0.78& 0.81& 0.79& 0.73 && 0.85 \\
\textbf{BI} && 0.54& 0.58& 0.57& 0.58 && 0.42 \\
\end{tabular}
\end{table}

To fully answer \textbf{RQ2} and specifically to verify H1 and H2 hypothesis, we performed a full factor analysis through structural equation modelling. All the analyses explained in the following were conducted using R with the Lavaan package version 0.6-8.

First, we conducted an Explanatory Factor Analysis (EFA)~\cite{EFA}, which revealed that the number of factors calculated graphically and analytically with, respectively, the methods Optimal Coordinates, parallel analysis, and Kaiser rule is equal to 3. Thus no other underlying latent variables emerged from the data. This analysis confirmed our initial hypothesis (represented in Figure~\ref{fig:model}) of 3 factors representing an overall intrinsic motivation construct (grouping together interest, value, trust and relatedness), behavioural intention and awareness creation.

Then, we conducted a  Confirmatory Factor Analysis (CFA)~\cite{CFA} to test H1 and H2. 
We evaluated the goodness of fit using the most commonly used indexes, in particular the comparative fit index (CFI), the Tucker Lewis Index (TLI), the standardised root mean square residual (SRMR) and the root mean square error of approximation (RMSEA). 

\begin{figure}[ht]
    \centering
    \includegraphics[width=0.7\textwidth]{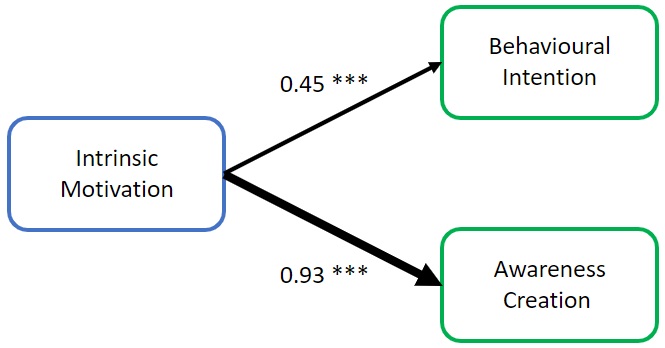}
    \caption{Model regression values and significance level (*** means $\alpha < 0.001$).}
    \label{fig:modelwithweights}
\end{figure}

The result of the CFA are represented in Figure~\ref{fig:modelwithweights} and confirm both our hypotheses. Indeed, the data shows that intrinsic motivation has a statistically significant and quite strong positive effect on both awareness creation (0.93) and behavioural intention (0.45). The model shows good or very good statistical fit measures (CFI 0.91, TLI 0.90, SRMR 0.057 and RMSEA 0.1) and a pretty high explained variance (approximately 63.3\%, which also may be a reason for RMSEA $>$ 0.8).

\begin{figure}[h]
    \centering
    \includegraphics[width=0.7\textwidth]{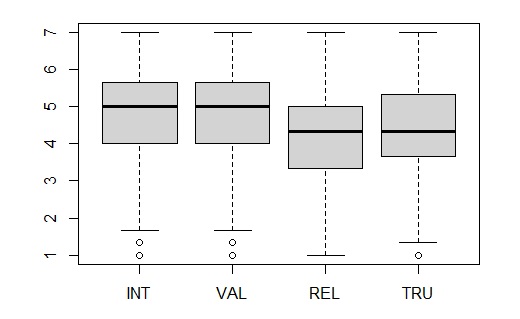}
    \caption{Boxplot of the Intrinsic Motivation sub-factors}
    \label{fig:boxplotIMI}
\end{figure}

Even if the factor analysis did not allow us to distinguish the contribution of the intrinsic motivation sub-factors, we had a closer look to their respective scores. The answer distribution of the items representing interest, value, trust and relatedness are displayed in the boxplot in Figure~\ref{fig:boxplotIMI}. All the sub-factors show quite high values in the 1-7 range: interest has mean 4.71, median 5, std dev 1.31; value has mean 4.62, median 5, std dev 1.42; trust has mean 4.39, median 4.33, std dev 1.23; and finally relatedness has mean 4.07, median 4.33, std dev 1.42. A t-test confirmed that there is no statistically significant difference between the sub-factors, even if the slightly lower scores on trust and relatedness may indicate that those are the aspects on which more attention should be given when designing a tool to solicit awareness creation and behavioural change.

All in all, we can therefore say that it is clear that intrinsic motivation factors influence both awareness creation and behavioural intention. Further research may be needed to distinguish the weight and influence of more specific dimensions; for example, the inclusion of conversational AI technology (opposed to the "scripted" conversational agent that we employed in our experiments) may indeed have a possibly negative effect on user trust and overall behavioural intention, with its possible misunderstanding w.r.t. user responses and intentions.

\subsection{Effect of conversational style}

Our third research question aims at evaluating whether changing the linguistic and communication style influences the users' behavioural intention. As explained in the methodology section, we included this aspect as experimental condition that distinguishes the three different campaigns.

\begin{figure}[h]
    \centering
    \includegraphics[width=0.6\textwidth]{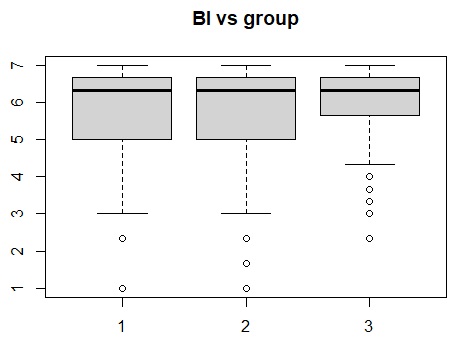}
    \caption{Boxplot of the intention to use a specific conversational style (1 informal, 2 informal with multimedia elements, 3 formal)}
    \label{fig:boxplotBIvsGroup}
\end{figure}

To answer \textbf{RQ3}, first we performed a pairwise t-test to compare the behavioural intention scores across the three experimental conditions (informal style, informal style with multimedia elements, formal style), whose distribution is displayed in Figure~\ref{fig:boxplotBIvsGroup}: group 1 has mean 5.84, median 6.33, std dev 1.17; group 2 has mean 5.69, median 6.33, std dev 1.36; and group 3 has mean 5.85, median 6.33, std dev 1.16. No statistically significant difference resulted from the t-test.

We also performed a group analysis, using the conversational style as grouping dimension in the confirmatory factor analysis (CFA), to see if any difference emerged between the groups when fitting the model. Also in this case, the data did not show any significant effect of the conversational style on the model fit.

Therefore, we can conclude that our experimental results do not support the hypothesis that the conversational style (in terms of language and communication approach) change the propensity of the user to adopt a conversational tool to support his/her well-being. This result may be due to the relatively small differences between the experimental conditions (i.e., the two informal style versions differed only for the insertion of emoticons and GIFs, and indeed their score distribution are virtually identical as displayed in Figure~\ref{fig:boxplotBIvsGroup}) or may be caused by the quite short conversational experience included in our experiment (mean conversation chatting time 2 mins and 51 secs) which may not be representative of the long-term effect that a repeated use of a conversational coaching tool may have. In this regards, more detailed work can be performed to evaluate the potential role of the conversational style.

\subsection{Content analysis}

The following is the content analysis conducted to answer our fourth research question \textbf{RQ4}, which aims to test the presence of participants' behavioural changes according to the topic of the conversation with the counsellor.

As previously reported, conversational experience covers some main aspects of well-being and healthy lifestyle, corresponding, respectively to \emph{self-care}, \emph{social relationships}, \emph{mental health} and \emph{healthy eating}. During the conversation, the participant can choose between two branches: "self-care and social relationships" and "mental health and healthy eating". Therefore, according to the choice, the topics of interaction between the participant and the counsellor differ. From an analysis carried out over the total of participants to all the experimental campaigns, the results show that 68\% of them prefer to discuss mental health and a healthy diet, while the remaining 32\% are interested in self-care and social aspects.

Finally, the topic chosen at the end of the interaction to improve the user lifestyle (the "promise" of the Will part of the GROW coaching model) was analysed and compared to the branch selected. The proposed topics were  \emph{healthy eating}, \emph{physical activity}, \emph{social relationships} and \emph{mental health}. For both the branches, \emph{physical activity} was the most chosen option (56.4\% of the participants in the first branch on "mental and healthy eating", and 48.5\% in "self-care and social relations" branch); this may be due to the popularity of the topic and the relatively easy activity that the user had to promise doing ("having 30 minutes of fast walking at least twice a week and avoid being too sedentary"). The second most frequent answer for the first branch was the proposal on \emph{healthy eating}, which reached 25.1\% of the preferences ("eat fruit or vegetables every day and avoid fatty, low-nutrient foods") and, for the other branch, \emph{social relationships} (25.7\%, "talk every day with people who make you feel good and who motivate you to achieve your goals"). 

Of the four subgroups with a choice in the four topics, only \emph{healthy eating} showed a significant statistic difference at the promise score between the two branches (t(15.361)=2.6243, p-value=0.01887, the mean in the first branch that included healthy eating topic is 6.39, while mean in the second branch is 4.86), which means that in general there was no great influence of the initial conversation topic on the actual promise for behavioural change. This may be due to the short conversation experience as well as to the fact that the crowdsourcing users just simulated a coaching experience (instead of actively participating in a broad and long-term coaching program).

\section{Conclusions and Next Steps}\label{sec:concl}

In this study, we proposed a model to analyse an individual's intention to adopt a conversational digital tool presented as a well-being counsellor, in an awareness creation and behavioural change setting, by comparing a set of motivation factors and communication styles that can influence awareness and solicit possible lifestyle changes. 

The results of our study confirm that our digital counsellor proved to be effective for the involved participants; our model also showed that the intrinsic motivation factors positively influence both awareness creation and behavioural intention; anyway, more deep studies might be needed to distinguish different sub-factors like trust and relatedness. On the other hand, our experimental data did not show any significant difference between the different conversational styles; in this regards, longitudinal studies in which participants repeatedly interact with the conversational counsellor may be more effective in eliciting such differences (if any).

We also carried out a content analysis about the choices by participants in the experimental campaign concerning the proposed health issues (mental health, healthy eating, self-care and social relationships). Finally, we collected all the open comments about the positive and negative aspects of the Allegra counsellor, and we performed a sentiment analysis. The gathered opinions are wide and vary, but report, in significant numbers, positive and quite positive comments that indicate Allegra as an easy-to-use, immediate, simple and clear tool. The informative messages offered to the user in the health domain are useful, and the friendly and non-judgmental tone is also much appreciated. Some critical aspects emerging from these analyses concern the level of information detail that some users, who are already familiar with the topic, already know, and the actual lack of personalisation.

Our next steps will be focused on extending the interaction of the user with the conversational tool: indeed, we think that a more long-term iteration of the coaching pattern (i.e., multiple short conversations, performed during several days, e.g. up to one month), even if carried out on a smaller number of participants, could be useful both to better monitor the real intention to behavioural change and the long-term effect of different communication styles. 

It is also worth noting that, in order to have a consistent system behaviour and equal experimental conditions among the participants, we based our work on a conversational survey instrument which provide full-scripted conversations (i.e., the interaction between the user and the conversational counsellor was identical for all participants to each of the experimental campaigns). Adopting conversational AI and more autonomous chatbots may have a different effect on intrinsic motivation, awareness creation and behavioural intention, because of the variability and partial unpredictability of the conversational agent. 

\section*{Acknowledgments}
This research was performed in the context of the DIPPS project (Digital Intervention in Psychiatric and Psychologist Services), partially co-funded by the Italian Ministry of Enterprises and Made in Italy (MIMIT).

\end{document}